\documentclass[twocolumn]{aastex631}

\shorttitle{Stability and Dynamics of the Proxima Centauri System}
\shortauthors{Livesey et al.}

\graphicspath{{./}{}}

\usepackage{physics}
\usepackage{hyperref}

\def\msun{{M_\odot}}

\def\mearth{{M_\oplus}}
\def\rearth{{R_\oplus}}
\newcommand\vplanet{{\footnotesize\texttt{VPLanet}}}
\newcommand\eqtide{{\footnotesize\texttt{EqTide}}}
\newcommand\distorb{{\footnotesize\texttt{DistOrb}}}
\newcommand\ecapsv{{\footnotesize\texttt{VSPACE}}}
\newcommand\rebound{{\footnotesize\texttt{REBOUND}}}
\newcommand\whfast{{\footnotesize\texttt{WHFAST}}}
\newcommand\iasfifteen{{\footnotesize\texttt{IAS15}}}

\newcommand{\revone}[1]{#1}
\newcommand{\revtwo}[1]{#1}

\DeclareMathOperator{\sgn}{sgn}

\begin{document}

\title{Orbital Stability and Secular Dynamics of the Proxima Centauri Planetary System}

\correspondingauthor{Joseph R. Livesey}
\email{jrlivesey@wisc.edu}

\author[0000-0003-3888-3753]{Joseph R. Livesey}
    \affiliation{Department of Astronomy, University of Washington, Seattle, WA 98195}
    \affiliation{Department of Astronomy, University of Wisconsin--Madison, Madison, WI 53706}

\author[0000-0001-6487-5445]{Rory Barnes}
    \affiliation{Department of Astronomy, University of Washington, Seattle, WA 98195}
    \affiliation{NASA Virtual Planetary Laboratory Lead Team, USA}

\author[0000-0001-9423-8121]{Russell Deitrick}
    \affiliation{School of Earth and Ocean Sciences, University of Victoria, Victoria, BC V8P 3E6}

\begin{abstract}
The two innermost planets of the Proxima Centauri system are separated by just 0.02 au, inducing strong gravitational interactions between them. We assess this interaction by leveraging fast orbital stability indicators and find that orbital stability is very likely if the initial eccentricities of planets b and d are less than $\sim0.2$, but cannot confirm stability at larger values. We find that stability is not strongly affected by the true masses of the planets or by the distant planet c. However, mutual inclinations between $95^\circ$ and $142^\circ$ often result in unstable motion. \revone{We further explore the long-term evolution of the orbits in these stable regions of parameter space and find that circularization can take over 5 Gyr. This tidal evolution could support surface energy fluxes in excess of 1 W m$^{-2}$} \revtwo{for over 1 Gyr, possibly affecting planet b's habitability.} 
\end{abstract}



\section{Introduction}
Proxima Centauri is the nearest stellar neighbor to the Sun and is known to host at least 3 planets, including one $\sim$Earth-mass planet in the habitable zone \citep{Anglada-Escude2016, Damasso2020, SuarezMascareno2020, Faria2022}. Given the system's proximity and the possible habitability of the middle planet, b, this system is uniquely valuable in the study of exoplanet formation, evolution, and habitability. Herein we describe a series of experiments to test the orbital stability of the Proxima Centauri system over a range of plausible orbital architectures.
 
The orbital evolution of Proxima's planetary system could affect the habitability of Proxima b in many ways. Orbital oscillations could affect the seasons and trigger snowball states \citep{Spiegel2009, Armstrong2014, Deitrick2018b, Wilhelm21}, while tidal damping can drive long term decay of orbital semi-major axis and eccentricity \citep{Barnes2008, Barnes2017}. Additionally, tidal friction in Proxima b's interior could \revone{create} large amounts of energy that affect the generation of magnetic fields \citep{DriscollBarnes2015}, vigorous volcanic activity \citep{Jackson2008c, Barnes2009}, and possibly even drive a runaway greenhouse \citep{Barnes2013}. Thus, defining the physically plausible range of orbital architectures in this system could significantly improve our understanding of Proxima b's habitability.

Not surprisingly, then, this study is not the first to consider the stability of this system. In particular, \cite{Meng2018} estimated the parameter space that permits orbital stability, but they used an older architecture that included only planets b and c. In this study, we update their results by considering the currently observed planetary system.

In particular, the recently detected planet d \citep{Faria2022} demands a reanalysis of this system's stability. The orbits of Proxima d and b are sufficiently near so that gravitational perturbations between the two could drive the two into instability, i.e., the ejection of a planet from the system. Such cases should therefore be deemed implausible starting conditions for the system, i.e., evolutionary models can safely avoid those unstable parameter combinations. Using semi-analytic and $N$-body simulations, we systematically searched for unstable configurations of this system. We also identify architectures that give rise to chaotic dynamics, which do not necessarily imply instability. \revone{Identifying these chaotic configurations is useful for investigating long-term behavior, as the dynamics of systems exhibiting orbital chaos are ill-described by a disturbing function containing only secular terms \citep[e.g., that employed in \distorb; see \textsection\ref{sec:secular} and][]{Barnes2020}.}

In \textsection\ref{sec:theory}, we outline the theoretical models we use to assess stability and the parameter space we will explore. In \textsection\ref{sec:results}, we present stability maps of the system obtained through thousands of simulations varying the initial conditions. In \textsection\ref{sec:discussion}, we discuss the major outcomes of this work and implications for future studies of this planetary system. Note that code for reproducing the results presented in this paper is available online.\footnote{\href{https://github.com/jrlivesey/ProximaStability}{https://github.com/jrlivesey/ProximaStability}} 

\section{Methods} \label{sec:theory}

In this section we describe theories of orbital stability, the numerical methods we employed to identify unstable cases, and the range of initial conditions we explored. Throughout this study we consider only the three-body system consisting of the host star and planets d and b. We tested the role of planet c, with an orbital period of $\sim$ 1,900 days, and found that it has negligible impact on the stability of the system, so we will not present detailed results of its role.

Proxima Centauri is a member of the $\alpha$ Centauri triple star system, orbiting the binary $\alpha$ Centauri AB at a distance of about 8700 au \citep{Kervella2017}. Asteroseismological analysis has revealed that the binary is $5.3 \pm 0.3$ Gyr old \citep{Joyce2018}. We assume that Proxima Centauri formed in the same cluster as $\alpha$ Centauri AB and therefore has the same age. While Proxima's planetary system could be modified by the binary \citep{Barnes2016}, we will ignore that possibility in our studies.

\subsection{\revone{Stability Criteria}} \label{sec:hill}
\revone{In this study we evaluate the Lagrange stability of the Proxima system. In a \textit{Lagrange stable} planetary system, all planets remain gravitationally bound to the star and the ordering of the planets is preserved. There is no analytic formula to determine whether a system is Lagrange stable, and so we resort to $N$-body simulations.}

We first consider, however, the possibility that the three-body system is Hill stable \citep{Szebehely1977, Marchal1982, Gladman1993, BarnesGreenberg2006}, which is determined analytically, but can only approximate orbital stability. The definition of \textit{Hill stability} is that the ordering of the bodies cannot change. In other words, if the outer planet escapes to infinity, the system would still be Hill stable. Although this definition is very limited, it does reduce to a single equation that can be quickly evaluated. \revone{In the present study, we assess the agreement between the ``Hill stability boundary'' of the system in orbital element space and the stability boundary obtained through simulations. Such an analysis has been performed previously for systems of two $\sim$Jupiter-mass planets \citep{BarnesGreenberg2006, BarnesGreenberg2007}, but not, to our knowledge, for systems of terrestrial planets.}

A system is guaranteed to be Hill stable if it satisfies the inequality $\beta > \beta_\text{crit}$ for
\begin{equation}
    \label{eq:exact_hill_stability_1}
    \beta = -\frac{2M}{G^2 \mathcal{M}^3} C L^2
\end{equation}
and
\begin{equation}
    \label{eq:exact_hill_stability_2}
    \beta_\text{crit} \simeq 1 + \frac{3^{4/3} m_1 m_2}{m_\star^{2/3} (m_1 + m_2)^{4/3}} - \frac{m_1 m_2 (11m_1 + 7m_2)}{3m_\star (m_1 + m_2)^2},
\end{equation}
where $m_\star$, $m_1$, and $m_2$ are the masses of the star, inner planet, and outer planet, respectively; $M = \sum_i m_i$ is the total mass of the system; $\mathcal{M} = (1/2) \sum_{i \neq j} m_i m_j$; $C$ is the total orbital energy of the system; and $L$ is its total angular momentum \citep{Marchal1982}. A three-body system for which $\beta/\beta_\text{crit} < 1$ is likely but not certain to be Hill unstable.

To evaluate Hill stability, we generated $10^6$ possible orbital configurations and calculated $\beta/\beta_\text{crit}$. We randomly sampled the longitudes of ascending node, arguments of periastron, and mean anomalies of both planets from $\mathcal{U}[0, 360)^\circ$. The inclinations of the planets were sampled from $\mathcal{U}[0, 15]^\circ$. The inclination of the planets relative to the sky plane, used to calculate their true masses, was sampled from $\mathcal{U}[10, 90] ^\circ$. For both of the orbital eccentricities, we used $e \sim \mathcal{U}[0, 0.9]$. We assumed the orbits were coplanar for these calculations.

\subsection{MEGNO}

A hybrid approach to estimating orbital stability is to compute a quantity called the MEGNO \citep[Mean Exponential Growth of Nearby Orbits;][]{Cincotta2000}, which measures the divergence of two trajectories in phase space. If $\boldsymbol{\delta}(t)$ denotes the separation between two initially infinitesimally close particles in phase space, then the MEGNO $Y$ is defined as
\begin{equation} \label{eq:megno}
    Y(t) = \frac{2}{t - t_0} \int_{t_0}^t \frac{\dot{\vb*{\delta}} \vdot \vb*{\delta}}{\vb*{\delta} \vdot \vb*{\delta}} t' \: \dd t'
\end{equation}
and its time-averaged value \revone{as}
\begin{equation} \label{eq:avg_megno}
    \langle Y \rangle (t) = \frac{1}{t - t_0} \int_{t_0}^t Y(t') \: \dd t'.
\end{equation}
For periodic (stable) evolution, $\langle Y \rangle \to 2$ as $t \to \infty$. \revone{For a chaotic system, $\langle Y \rangle \to (\gamma/2) t$, where $\gamma$ is the maximum Lyapunov characteristic number} \citep{Cincotta2000, Gozdziewski2001, Morbidelli2002, Cincotta2003, Hinse2010}. The MEGNO is defined similarly to the Lyapunov characteristic number, but is weighted with time, amplifying stochastic behavior in the evolution and allowing for an earlier detection of chaos \citep{Cincotta2000, Gozdziewski2001, Morbidelli2002}. We follow standard practice, and label the motion as chaotic if $| \langle Y \rangle - 2 | > 10^{-2}$, indicating the system may be unstable on Gyr timescales.

\subsection{General Relativistic Effects}
Relativistic (GR) effects can modify the precessional frequency of the orbit by an amount
\begin{equation}
    \dot{\varpi}_\text{GR} \simeq \frac{3G m_\star n}{c^2 a (1 - e^2)} = \frac{3(G m_\star)^{3/2}}{c^2 a^{5/2} (1 - e^2)},
\end{equation}
where $n$ is the mean motion of the planet in question \citep{Sterne1939}. If we apply this relation to the most extreme case studied here, i.e., that with $e = 0.9$ for both planets,  we obtain GR precession periods of about 22,000 years and 82,000 years for planets d planet b, respectively. \revone{Since previous work has shown that $N$-body integrations up to $10^6$ orbits of the outermost body are sufficient to identify most unstable configurations of a planetary system \citep{Barnes2004}.} We therefore run our simulations for about $10^6$ orbits of planet b, which is $\sim$ 31,000 yr. Therefore, GR will only induce one extra circulation over the length of our simulation and should therefore not be a significant effect. To decrease computational cost, we therefore do not include a post-Newtonian correction to our model.

By the same token, we may safely assume that the rate of precession due to the Lense--Thirring effect is also negligible \citep{Will1972, Iorio2008}.

\subsection{\revone{N-Body Simulations}}
Table \ref{tab:obs-params} lists the current best fits and uncertainties for the physical and orbital elements of the system. We assume the host star has a mass of $0.12\msun$ \citep{Kervella2017}. However, we are also interested in the range of plausible architectures at the end of planet formation as tidal damping can cause long-term changes to the planetary orbits \citep[e.g.,][]{Barnes2008, Barnes2017, Meng2018}. We will thus consider a wider parameter space than allowed by observational uncertainties, focusing on larger eccentricities and semi-major axes.

\begin{deluxetable*}{c|cccc}
\tablecaption{Best-fit parameters for the Proxima Centauri system.\label{tab:obs-params}}
\tablewidth{0pt}
\tablehead{
\colhead{Body} & $a$ & $e$ & \colhead{$m \sin i_\text{obs}$} & $i_\text{obs}$
\\
\colhead{} & \colhead{(au)} & \colhead{} & \colhead{} & \colhead{(deg)}
}
\startdata
star\tablenotemark{a} & & & $0.1221 \pm 0.0022 \msun$ & \\
Planet d\tablenotemark{b} & $0.02885^{+0.00019}_{-0.00022}$ & $0.04^{+0.15}_{-0.04}$ & $0.26 \pm 0.05 \mearth$ & \nodata \\
Planet b\tablenotemark{b} & $0.04856^{+0.00030}_{-0.00030}$ & $0.02^{+0.04}_{-0.02}$ & $1.07 \pm 0.06 \mearth$ & \nodata \\
Planet c\tablenotemark{c} & $1.48^{+0.08}_{-0.08}$ & $0.04^{+0.01}_{-0.01}$ & $5.8 \pm 1.9 \mearth$ & $133 \pm 1$\tablenotemark{d}
\enddata
\tablenotetext{a}{\citet{Kervella2017}}
\tablenotetext{b}{\citet[][TV RM analysis]{Faria2022}}
\tablenotetext{c}{\citet{Damasso2020}}
\tablenotetext{d}{\citet{Benedict2020b}}
\end{deluxetable*}

To gauge the dependence of the system's stability on the system architecture, we performed three sets of 10,000 $N$-body simulations with the publicly available \rebound\ code\footnote{Available at \href{https://github.com/hannorein/rebound}{https://github.com/hannorein/rebound}. Our simulations use v3.28.3.} \citep{Rein2012, Rein2015, Rein2015b, Rein2016}. In Set I we varied the initial eccentricities of both planets; in Set II we varied the planets' masses; and in Set III we varied the initial inclinations with respect to the fundamental plane.

\revone{In addition to evaluating the MEGNO, in our Set I simulations we identify the parameter space in which the orbits cross, and in which we therefore anticipate chaos. As there are two planets included in our $N$-body simulations, planet d's apocenter coincides with planet b's pericenter, and thus the orbits cross where $a_b (1 - e_b) < a_d (1 + e_d)$. We also incorporate the two-planet chaos criterion given by \citet{HaddenLithwick2018} in terms of the relative eccentricity, which in our case simplifies somewhat to}
\begin{align}
    e_d + e_b &> \left ( \frac{a_b - a_d}{a_d} \right ) \nonumber \\
    &\quad \times \exp \left [ -\frac{11}{5} \left ( \frac{m_d + m_b}{m_\star} \right )^{1/3} \left ( \frac{a_b}{a_b - a_d} \right )^{4/3} \right ].
\end{align}

\revone{In our Set II simulations, we test the stability of the system with combinations of masses up to $10 \mearth$ for both planets. Previous studies have specified a critical orbital separation in a two-planet system with low $e$ and $i$, at which the system becomes unstable. The critical separation is $\Delta a = 2\sqrt{3} R_\text{H}$, where
\begin{equation} \label{eq:hill-radius}
    R_\text{H} = \frac{a + a'}{2} \left ( \frac{m + m'}{3m_\star} \right )^{1/3}
\end{equation}
is the Hill radius \citep{HasegawaNakazawa1990, Gladman1993, Chambers1996}. Primed quantities refer to the outer body. In the most extreme case considered here, both masses are $10 \mearth$, with which $\Delta a = 36.3 R_\text{H}$. We therefore expect every experiment in Set II to result in stable motion; we perform them and include the results here for completeness.}

The \textit{mutual inclination} (or relative inclination) $\Psi$ between the two planets' orbits is given by
\begin{equation} \label{eq:mutualinc}
    \cos(\Psi) = \sin(i) \sin(i') \cos(\Omega - \Omega') + \cos(i) \cos(i').
\end{equation}
\revone{Note that because there are only two planets included in our $N$-body simulations, $\cos(\Omega - \Omega') \simeq -1$ for our purposes.} The resulting distribution of simulated cases is weighted toward higher values of $\Psi$, since nearly coplanar, prograde orbits are already expected to be stable.

\revtwo{For each set of simulations, we identify the Hill stability boundary in the relevant parameter space as an additional analytic stability criterion. We compute this boundary for each set because the Hill stability parameter depends upon the eccentricities, masses, and inclinations of both planets according to Eq.~(\ref{eq:exact_hill_stability_1}).}

The simulations are integrated for $\sim 10^6$ orbits of Proxima b. This \revtwo{timescale} is longer than the conventional integration times used for identifying chaotic motion with the MEGNO\revtwo{, which} are typically $\sim 10^{3-4}$ times the longest dynamical timescale in the system \citep{Cincotta2000, Hinse2010, Meng2018}. We perform our integrations over this longer timescale because we find some cases that are surprisingly stable (see Section \ref{sec:orbital-stability}).

The initial conditions for all of our \rebound\ simulations are provided in Table \ref{tab:initial-conditions}. The semi-major axes used are slightly greater than \revone{the} measured values to account for tidal effects over time. The approximate mass values used in these simulations for both planets were obtained by setting $i_\text{obs} = 133^\circ$, the mean observational inclination of Proxima c obtained astrometrically by \citet{Benedict2020b}.

\begin{deluxetable*}{lc|cccccc}
\tablecaption{Initial conditions for the \rebound\ simulations.\label{tab:initial-conditions}}
\tablewidth{0pt}
\tablehead{
\colhead{Set} & \colhead{Planet} & \colhead{$a$} & \colhead{$e$} & \colhead{$m \sin i_\text{obs}$} & \colhead{$i$} & \colhead{$\Omega$} & \colhead{$\varpi$} \\
\colhead{} & \colhead{} & \colhead{(au)} & \colhead{} & \colhead{($\mearth$)} & \colhead{(deg)} & \colhead{(deg)} & \colhead{(deg)}
}
\startdata
I & d & 0.029 & 0--0.9 & 0.26 & 0 & 0 & 0 \\
 & b & 0.049 & 0--0.9 & 1.07 & 0 & 180 & 180 \\
\tableline
II & d & 0.029 & 0 & 0.21--10 & 0 & 0 & 0 \\
& b & 0.049 & 0 & 1.01--10 & 0 & 180 & 180 \\
\tableline
III & d & 0.029 & 0 & 0.26 & 0--180 & 0 & 0 \\
& b & 0.049 & 0 & 1.07 & 0--180 & 180 & 180 \\
\enddata
\end{deluxetable*}

We performed two varieties of $N$-body simulations with \rebound. We first used \whfast, a symplectic Wisdom--Holman integrator \citep[see][]{Rein2015, Wisdom1991} to integrate the simulations. \whfast\ is fast and efficient, but does not fare well when the planets' orbits are strongly perturbed (i.e., when they become significantly non-Keplerian). In cases where the relative energy error exceeds $10^{-6}$, we re-run the simulations using \iasfifteen, a 15th order Gauss--Radau integrator \citep{Rein2015b}. \iasfifteen\ employs adaptive timestepping to resolve close encounters. At the end of each simulation, we find $\langle Y \rangle$ using built-in functions in \rebound\ that employ variational equations to calculate the trajectories of nearby orbits \citep{Rein2016}.

\revone{
\subsection{Secular Orbital Evolution} \label{sec:secular}
The $N$-body experiments discussed in the previous section yield a set of points in initial parameter space that give rise to quasi-periodic motion, where \revtwo{$\langle Y \rangle \to 2$}, and are thus suitable starting points for long-term simulations of the secular evolution.
}

\revone{
We performed six such simulations using the modular planetary evolution code \vplanet\footnote{Available at \href{https://github.com/VirtualPlanetaryLaboratory/vplanet}{https://github.com/VirtualPlanetaryLaboratory/ vplanet}. Our simulations use v2.3.28.} \citep{Barnes2020}. In particular, we used the \eqtide\ and \distorb\ modules to simultaneously simulate evolution of the orbits due to equilibrium stellar tides \citep[the ``constant phase lag'' model of][]{GoldreichSoter1966} and due to the disturbing function parameterizing gravitational interactions between the planets \citep{MurrayDermott1999}. For detailed descriptions and validation of these models, the reader is directed to \citet{Barnes2020} and \citet{Deitrick2018a}.
}

\revone{
Each body in the system is given a value of $k_2$ and $Q$. $k_2$ is the Love number of degree 2 and $Q$ is the ``tidal quality factor'' that parameterizes the rate of tidal dissipation in a body's interior. Earth has $Q = 12$ \citep{Williams1978}, but a terrestrial planet without large oceans may have a quality factor an order of magnitude larger. In each of our simulations, $k_2 = 0.5$ for the star, $k_2 = 0.3$ for the planets. The star's $Q$ is set to $10^6$, while the planet's $Q$ is sampled in the range $\mathcal{U}[10, 10^3]$. Planetary radii are necessary for calculating the tidal torques, and are determined according to the power law $R \propto m^{0.274}$ \citep{Sotin2007}.} \revtwo{We can only estimate these radii, as no planet in this system is transiting \citep{Jenkins2019}. The values we use fall within the range established by the examination of Proxima b's internal structure given in \citet{Brugger2016}.}

\revone{In addition to evaluating the time dependence of the orbital elements, we calculate the tidal heating of planet b throughout each evolution according to $\dot{E}_\text{tide} = -(\dot{E}_\text{orb} + \dot{E}_\text{rot})$, where}
\begin{equation} \label{eq:orb-tide-heat}
    \dot{E}_\text{orb} = \frac{Z}{8} \left [ 4\zeta - \left ( 20\zeta + 76 \right ) e^2 \right ]
\end{equation}
\revone{and}
\begin{equation} \label{eq:rot-tide-heat}
    \dot{E}_\text{rot} = \frac{Z}{8} \frac{\omega}{n} \left [ 4\zeta - \left ( 20\zeta + 48 \right ) e^2 \right ]
\end{equation}
\revone{are the power dissipated from the orbit and rotation of the planet, respectively.\footnote{Our Equations \ref{eq:orb-tide-heat} and \ref{eq:rot-tide-heat} are simplified from the CPL tidal heating equations in \citet[][see their Appendix E]{Barnes2020}. We set $\omega/n = 1$ at the start of each simulation, and assume that $\omega/n \simeq 1$ throughout. Thus, we take $\sgn(2\omega - 3n) = -1$, $\sgn(2\omega - n) = +1$ and $\sgn(n) = +1$. Proxima b has zero obliquity in each of our simulations, and so we discard terms related to obliquity tides.} Here, $n$ and $\omega$ are the orbital and rotational frequencies, $\zeta = \sgn(2\omega - 2n)$, and}
\begin{equation}
    Z = 3G^2 k_2 m_\star^2 (m_\star + m) \frac{R^5}{a^9} \frac{1}{n Q}.
\end{equation}

\revone{
Initial conditions were sampled randomly from the permitted region of parameter space using \ecapsv, a tool for initializing \vplanet's parameter sweeps. They are presented in Table \ref{tab:initial-conditions-vplanet}.} \revtwo{In these simulations, we include planet c and} \revone{assume for the sake of simplicity that Proxima c's orbit begins with zero inclination and does not evolve tidally. Therefore, values of \revtwo{$R$,} $\Omega$, and $Q$ are not provided for this planet. 
}

\begin{deluxetable}{ccc}
\tablecaption{Distributions from which the initial conditions of our \vplanet\ simulations are sampled.\label{tab:distributions-vplanet}}
\tablewidth{0pt}
\tablehead{
\colhead{Planet} & \colhead{Quantity} & \colhead{Distribution} \\
}
\startdata
d & $P_\text{orb}$ & $\mathcal{U}(5.122, 5.4)$ days \\
& $Q$ & $\mathcal{N}(100, 250)$ \\
& $\Omega$ & $\mathcal{U}(0, 360)$ deg \\
& $\omega$ & $\mathcal{U}(0, 360)$ deg \\
\tableline
b & $P_\text{orb}$ & $\mathcal{U}(11.187, 11.5)$ days \\
& $Q$ & $\mathcal{N}(100, 250)$ \\
& $\Omega$ & $\mathcal{U}(0, 360)$ deg \\
& $\omega$ & $\mathcal{U}(0, 360)$ deg \\
\tableline
c & $P_\text{orb}$ & $\mathcal{N}(1902, 24)$ days \\
& $\Omega$ & $\mathcal{U}(0, 360)$ deg \\
& $\omega$ & $\mathcal{U}(0, 360)$ deg \\
\enddata
\end{deluxetable}

\begin{deluxetable*}{cc|cccccccc}
\tablecaption{Initial conditions for the \vplanet\ simulations, numbered 1--6.\label{tab:initial-conditions-vplanet}}
\tablewidth{0pt}
\tablehead{
\colhead{Simulation} & \colhead{Planet} & \colhead{$P$} & \colhead{$e$} & \colhead{$m$} & \colhead{\revtwo{$R$}} & \colhead{$i$} & \colhead{$\Omega$} & \colhead{$\varpi$} & \colhead{$Q$} \\
\colhead{} & \colhead{} & \colhead{(days)} & \colhead{} & \colhead{($\mearth$)} & \colhead{\revtwo{($\rearth$)}} & \colhead{(deg)} & \colhead{(deg)} & \colhead{(deg)} & \colhead{}
}
\startdata
1 & d & 5.23736 & 0.12679 & 0.51653 & \revtwo{0.83443} & 5.35445 & 214.29919 & 301.19731 & 389.31695 \\
& b & 11.49553 & 0.13508 & 2.18208 & \revtwo{1.23837} & 8.52253 & 278.74258 & 287.98729 & 85.28869 \\
& c & 1,881.12305 & 0.01 & 6 & & 0 & & 375.5932 & \\
\tableline
2 & d & 5.1691 & 0.09071 & 0.39862 & \revtwo{0.777237} & 5.35445 & 273.62635 & 211.36366 & 322.54608 \\
& b & 11.3132 & 0.0925 & 1.71255 & \revtwo{1.158826} & 8.52253 & 308.83944 & 221.8431 & 466.03801 \\
& c & 1,881.91953 & 0.01 & 6 & & 0 &  & 71.65173 & \\
\tableline
3 & d & 5.24638 & 0.12754 & 0.35092 & \revtwo{0.75056} & 2.0301 & 30.27576 & 62.72857 & 250.40924 \\
& b & 11.27027 & 0.0711 & 1.23215 & \revtwo{1.05887} & 8.76222 & 53.66543 & 195.10852 & 220.5822 \\
& c & 1,885.90255 & 0.01 & 6 & & 0 &  & 136.16203 & \\
\tableline
4 & d & 5.34536 & 0.17036 & 0.442 & \revtwo{0.80} & 2.0301 & 7.32011 & 72.85377 & 277.98179 \\
& b & 11.32216 & 0.09628 & 1.13222 & \revtwo{1.03461} & 8.76222 & 130.82045 & 197.2157 & 646.02112 \\
& c & 1,914.12162 & 0.01 & 6 &  & 0 & & 360.19573 & \\
\tableline
5 & d & 5.37027 & 0.18371 & 0.33433 & \revtwo{0.74067} & 2.0301 & 283.00882 & 211.43098 & 243.22927 \\
& b & 11.4511 & 0.12222 & 1.23139 & \revtwo{1.05869} & 8.76222 & 264.70184 & 242.43391 & 15.94921 \\
& c & 1,892.59296 & 0.01 & 6 &  & 0 & & 329.46207 & \\
\tableline
6 & d & 5.35688 & 0.17726 & 0.32236 & \revtwo{0.73331} & 2.0301 & 62.43061 & 24.80976 & 452.43231 \\
& b & 11.33074 & 0.08986 & 1.37201 & \revtwo{1.09053} & 8.76222 & 174.8711 & 299.8717 & 124.37175 \\
& c & 1,866.94917 & 0.01 & 6 &  & 0 & & 185.42062 & \\
\enddata
\end{deluxetable*}

\section{Results} \label{sec:results}
\subsection{Orbital Stability} \label{sec:orbital-stability}

\begin{figure*}
    \centering
    \includegraphics{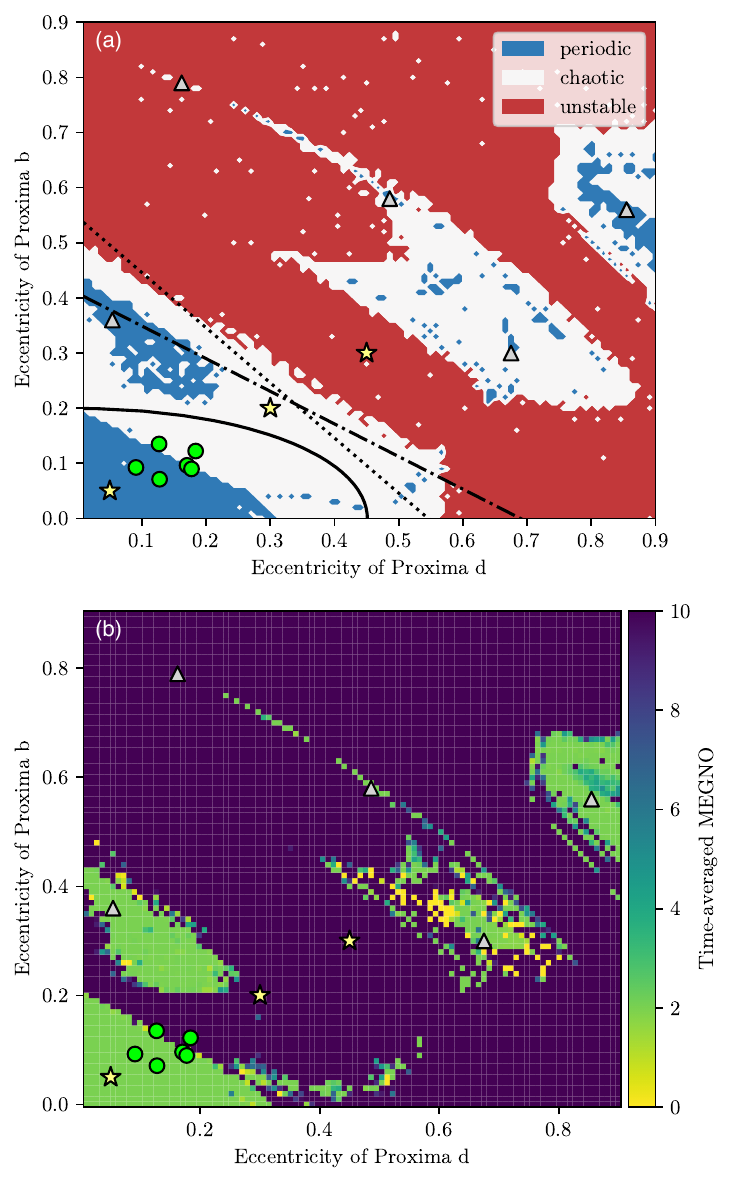}
    \caption{\revtwo{Panel (a) shows a} stability map of Proxima d and b, with both eccentricities varying between 0 and 0.9. Blue cells indicate periodic evolution according to the MEGNO criterion, red cells indicate one planet was ejected, and white cells indicate chaotic (but not necessarily unstable) evolution. The \revone{solid} black curve marks the Hill stability boundary in this plane, with Hill stable cases lying below the curve. \revone{The dashed-dotted black line indicates the limit of crossing orbits, i.e., above this line the apocenter distance of planet d is initially greater than the pericenter distance of planet b. The dotted line indicates the stability limit of \citet{HaddenLithwick2018}.} Yellow stars indicate the simulations shown in detail in Figure \ref{fig:all-ecc-evols}. \revone{Green dots indicate the locations of the secular dynamics simulations detailed in Section \ref{sec:secular}. Grey triangles indicate the locations of the longer (1 Myr) integrations described in Section \ref{sec:orbital-stability}. \revtwo{Panel (b) shows} a MEGNO map of the same parameter space. Most points in this map are similarly colored, as these simulations concluded with $\langle Y \rangle \geq 10$ or with an ejection, in which case an arbitrarily high MEGNO value was assigned.}}
    \label{fig:varying-ecc}
\end{figure*}

We find that the orbits of the two inner planets are Hill stable in just 7.2\% of the parameter space we explored. The Hill stable configurations are those with the lowest eccentricities for both planets and the boundary is shown with the black curve in Figure \ref{fig:varying-ecc}.

The results of Set I are also shown, with colors denoting values of $\langle Y \rangle$. Blue points indicate periodic (stable) configurations ($|\langle Y \rangle - 2| \leq 10^{-2}$), white points are chaotic ($|\langle Y \rangle - 2| > 10^{-2}$, but both planets remained bound), and red points are cases in which one of the planets escaped to infinity. A clear trend is present in which larger eccentricities lead to decreasing stability. \revone{There are many cases in which $\langle Y \rangle < 2$ --- or even becomes negative --- at the end of a simulation. Such trajectories lie in the neighborhood of stable periodic orbits \citep[e.g.,][]{Breiter2005}. There are cases in which the variational equations fail to converge at the end of a simulation, in which case \rebound\ returns no value for the MEGNO. This outcome means that the separation in phase space has grown very large, and thus we label these cases as chaotic without having an exact MEGNO value.}

We find that most periodic configurations lie within the Hill stability boundary, with $e_d \lesssim 0.25$ and $e_b \lesssim 0.2$. Most configurations with $e_d \lesssim 0.55$ and $e_b \lesssim 0.45$ give rise to motion that is chaotic, but does not become unstable within the length of these simulations. Beyond these boundaries, the system is very likely to be unstable, with one of the planets escaping to infinity within the length of these simulations.

We also find ``islands'' of configurations at higher initial eccentricities that are mostly classified as chaotic, but with some periodic results, e.g., near $(0.55, 0.4)$ and $(0.85, 0.55)$. The observational uncertainties suggest the system cannot currently exist in these Hill unstable islands (see Table \ref{tab:obs-params}). We return to this possibility in \textsection\ref{sec:discussion}.

The outlying stable regions shown in Figure \ref{fig:varying-ecc} could correspond to mean motion resonances, however, we find no evidence that MMRs are responsible for stabilizing these configurations. \revone{For a selection of simulated high-$e$ stable cases, we calculated the time evolution of resonant arguments, which are}
\begin{equation} \label{eq:resonant-argument}
    \phi = j_1 \lambda' + j_2 \lambda + j_3 \varpi' + j_4 \varpi
\end{equation}
\revone{for the $j_1 : j_2$ MMR, where the coefficients satisfy the d'Alembert relation $\sum_i j_i = 0$. We found no libration in $\phi$ up to $j_1 = 30$.} We also re-ran five cases for 1 Myr, and found that all cases initially designated as periodic were reclassified as chaotic (\revone{i.e., $\langle Y \rangle \not\to 2$}). We therefore conclude that these outlying islands are in fact unstable and that the MEGNO method was not able to identify instabilities over the first $10^6$ orbits. \revone{This conjecture is supported by the fact that these anomalous cases lie in the crossing orbits region of parameter space, and thus chaotic evolution is expected.} Note that simulating the system for its age would require at least 5,000 CPU hours per simulation with \whfast\ on a modern workstation, and so is currently intractable.

\begin{figure*}
    \centering
    \includegraphics[width=\linewidth]{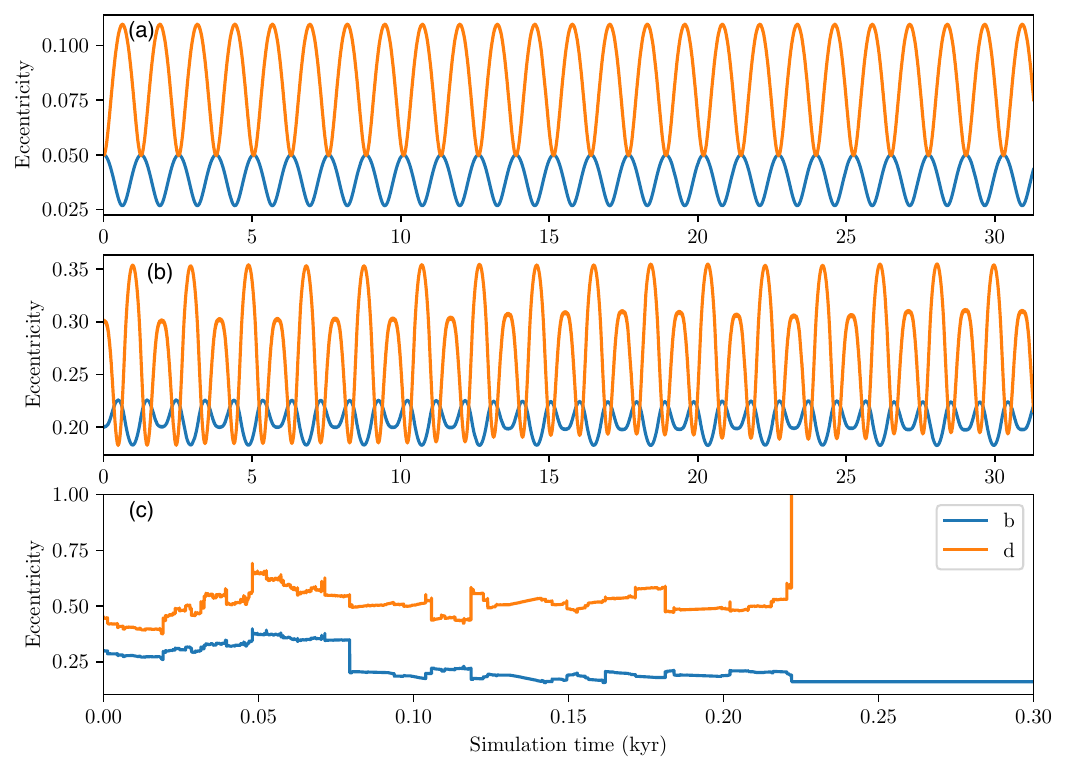}
    \caption{Example evolutions of Proxima d and b. Periodic evolution originating from the initial conditions $e_d = e_b = 0.05$ \revtwo{is shown in panel (a)}. An evolution originating from $e_d = 0.3$ and $e_b = 0.2$ \revtwo{is shown in panel (b)}. The time-averaged MEGNO for this case diverges significantly from 2, signifying that the motion is chaotic. At least 3 distinct frequencies are seen to be present in both planets' eccentricities, \revone{which is greater than the number predicted by secular theory}. An unstable case originating from $e_d = 0.45$ and $e_b = 0.3$ \revtwo{is shown in panel (c)}. The ejection of planet d from the system is evident at $t = 222$ yr, where $e_d > 1$.}
    \label{fig:all-ecc-evols}
\end{figure*}

Figure \ref{fig:all-ecc-evols} shows the eccentricity evolution for three example cases: a periodic case, a chaotic case, and an unstable case. The locations of these cases in Figure~\ref{fig:varying-ecc} are marked by stars. Recall that Set I assumes coplanar orbits, so there is no evolution with respect to the inclinations. The stable case shows classic periodic motion in which both eccentricities oscillate with just one frequency. The chaotic case appears regular, but contains at least two frequencies, probably resulting from the larger eccentricities. The unstable case ejects planet d in less than 250 years.

Next we turn to Set II, which focused on the role of planetary masses on stability. \revone{In each of these simulations, the initial orbits are circular and coplanar.} We found that every configuration we tested resulted in periodic motion. We conclude that stability is not strongly influenced by the planetary masses, which is consistent with the $m^{1/3}$ dependence in the Hill stability criterion.

\begin{figure*}
    \centering
    \includegraphics[width=\linewidth]{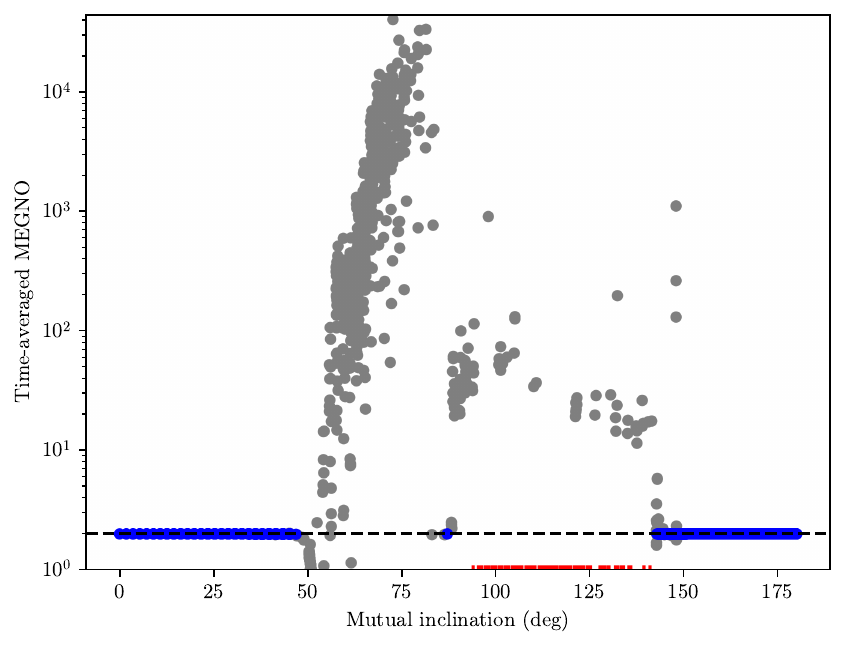}
    \caption{MEGNO values of the system over all possible initial mutual inclinations. The dashed black line indicates $\langle Y \rangle = 2$. Blue dots are cases with periodic or quasi-periodic motion, grey dots are cases with chaotic motion, and the red ticks indicate values of $\Psi$ for which the system is unstable (and therefore any MEGNO value is meaningless). All cases but one between 50 and $140^\circ$ resulted in chaotic motion. 165 simulations resulted in ejections, all with mutual inclinations between 95 and $142^\circ$.}
    \label{fig:varying-mutual-inc}
\end{figure*}

\begin{figure*}
    \centering
    \includegraphics{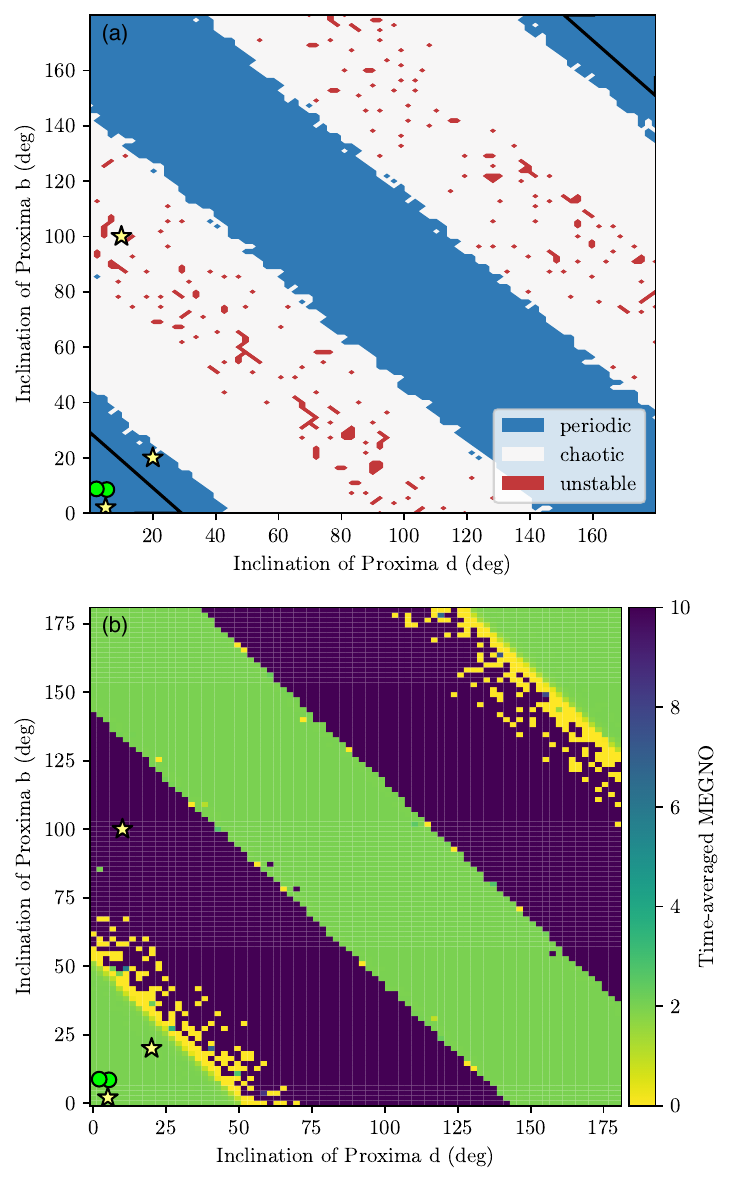}
    \caption{\revone{Similar to Figure \ref{fig:varying-ecc}, but with the initial inclinations varied instead of eccentricities. The black lines in \revtwo{panel (a)} indicate the Hill stability boundary. Yellow stars indicate the simulations presented in detail in Figure \ref{fig:all-inc-evols}, and the green dots point out two combinations of inclinations used in the secular dynamics simulations.}}
    \label{fig:varying-inc}
\end{figure*}

\begin{figure*}
    \centering
    \includegraphics[width=\linewidth]{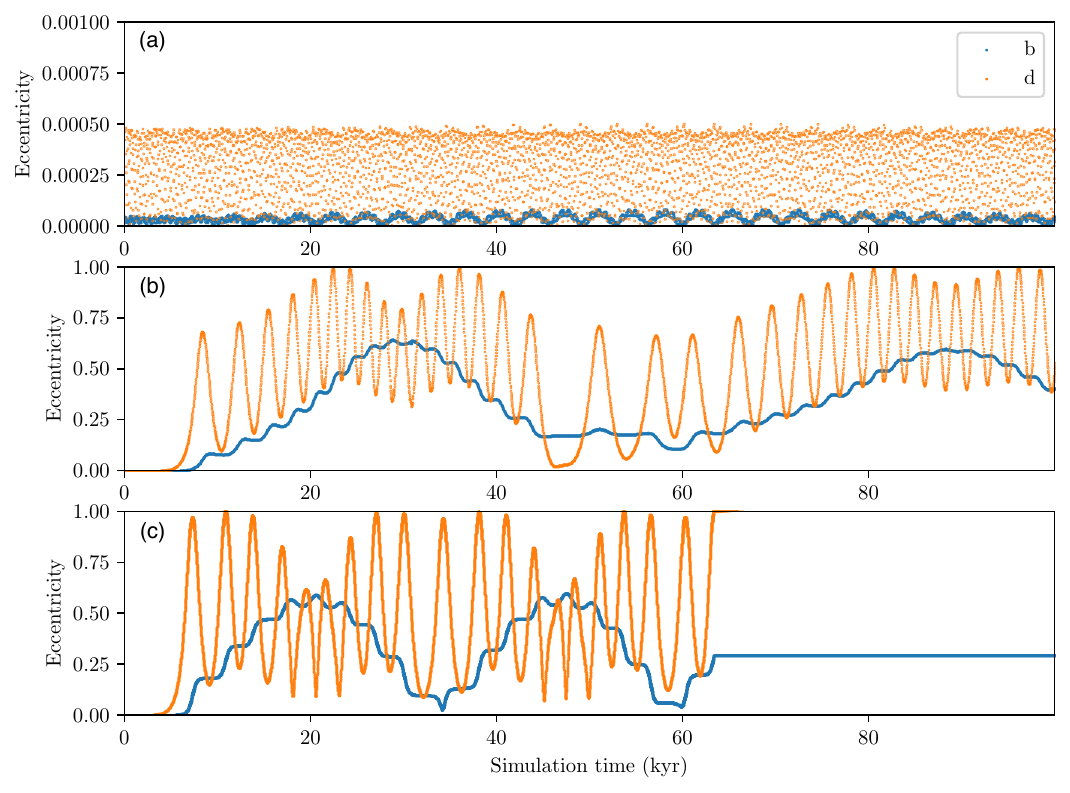}
    \caption{\revone{Sample evolutions of Proxima d and b, beginning with circular orbits. \revtwo{Panel (a) shows} periodic evolution, where initially $\Psi = 3^\circ$. \revtwo{Panel (b) shows} chaotic evolution, where initially $\Psi = 60^\circ$. \revtwo{Panel (c) shows} an unstable case originating from $\Psi = 110^\circ$. The ejection of planet d from the system is evident at $\sim 63$ kyr.}}
    \label{fig:all-inc-evols}
\end{figure*}

Finally, we present results for Set III, which focused on the inclinations. As shown in Figure \ref{fig:varying-mutual-inc}, we find that the motion is periodic for $\Psi < 50^\circ$ and $\Psi > 150^\circ$. With few exceptions, systems in between evolve chaotically. 80\% of systems with $95^\circ < \Psi < 142^\circ$ resulted in ejections. This finding is consistent with previous results that show stability increases \revtwo{as the orbits approach coplanarity} \citep[e.g.,][]{Meng2018}. \revone{Our obtained distribution of chaotic configurations in mutual inclination space, with a wide chaotic region in the middle skewed toward greater $\Psi$, agrees with the stability analyses of two-planet systems performed in \citet{VerasFord2010} in terms of the maximum eccentricity variation. Our Set III data are presented again in two dimensions --- $i_d$ and $i_b$ --- in Figure \ref{fig:varying-inc}. Figure \ref{fig:all-inc-evols} shows the eccentricity evolution for three example cases of various initial mutual inclination.}

\subsection{Long-term Orbital Evolution} \label{sec:long-term-evolution}

\begin{figure*}
    \centering
    \includegraphics[width=\linewidth]{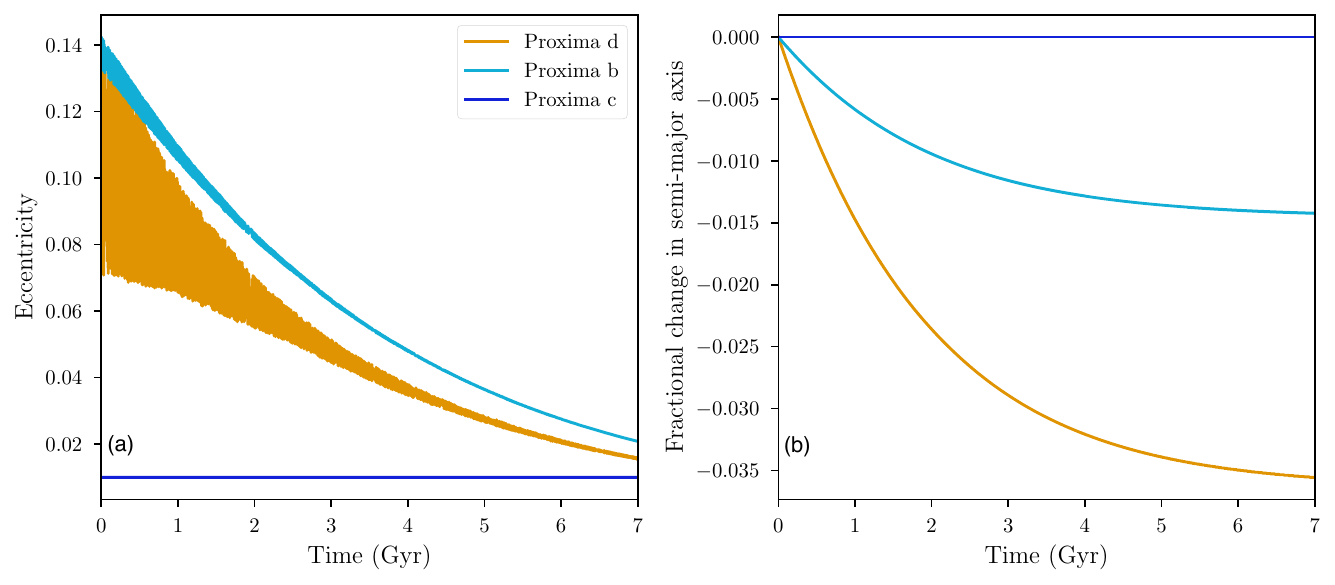}
    \caption{\revone{Secular orbital evolution of all three planets in Simulation 1. In this case, we use $e_d = 0.127$, $e_b = 0.135$, $m_d = 0.517 \mearth$, $m_b = 2.182 \mearth$, and $\Psi = 7.859^\circ$.}}
    \label{fig:secular-evol}
\end{figure*}

\revone{Using initial orbital configurations that give rise to regular motion, we simulated the long-term orbital evolution of the system. One result is shown in Figure \ref{fig:secular-evol}. Notably, in this scenario tidal evolution persists longer than the 7 Gyr for which the simulation was run, and much longer than the expected value for the age of the system.}

\begin{figure*}
    \centering
    \includegraphics[width=\linewidth]{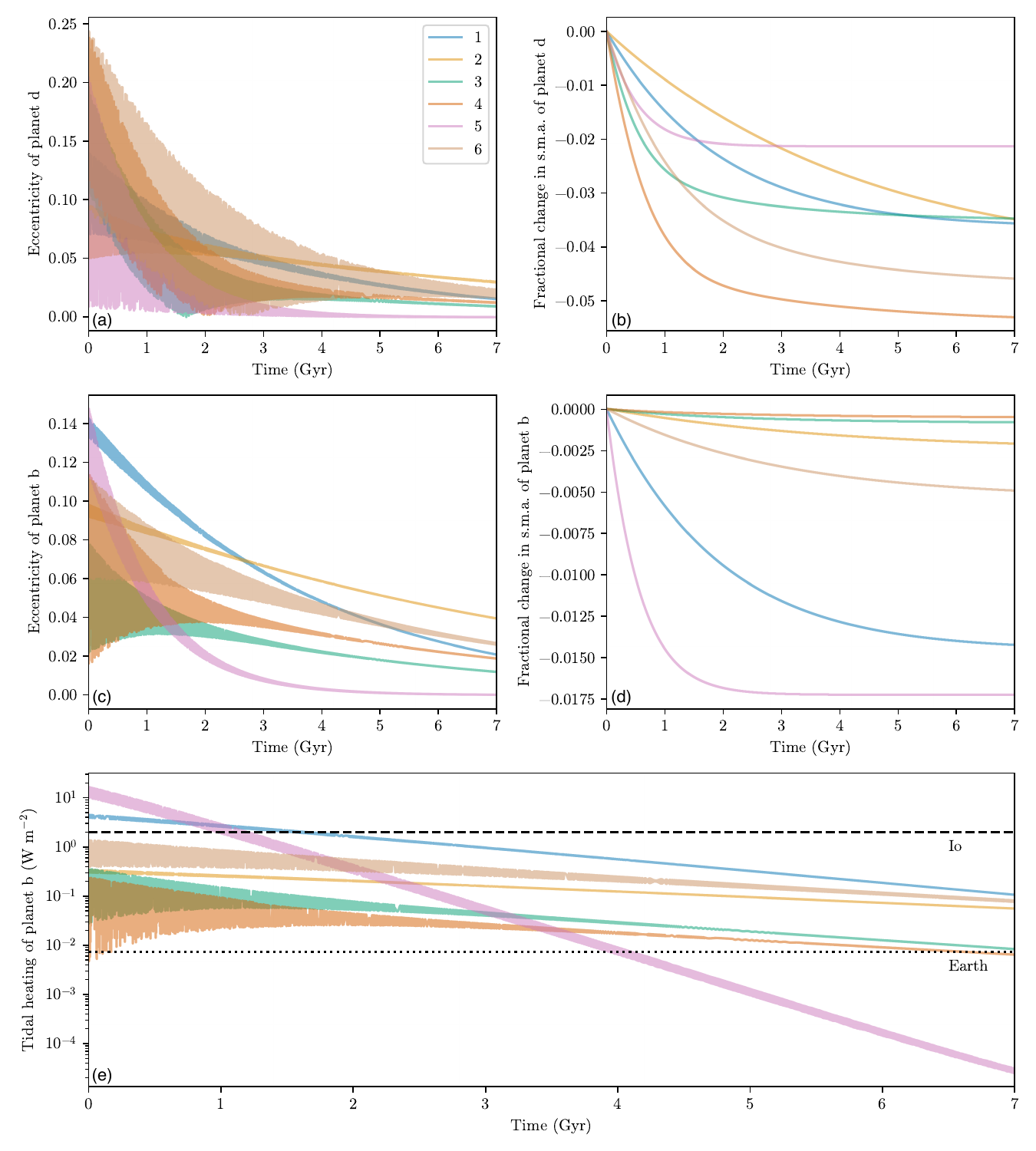}
    \caption{\revone{Secular orbital evolution of Proxima d in each simulated case \revtwo{is displayed in panels (a) and (b), and Proxima b in panels (c) and (d)}, with the tidal heat flux through the planet b's surface \revtwo{shown in panel (e)}. Three simulations end with a tidal heat flux of $\sim 0.1$ W m$^{-2}$, two end with a near-Earth value of $\sim 0.01$ W m$^{-2}$, and one with a far lower value. The present tidal heat flux at Io's and Earth's surfaces are indicated for reference \citep{Moore2007, MunkWunsch1998}.}}
    \label{fig:secular-evols}
\end{figure*}

\revone{The evolution of Proxima d and b in terms of $e$ and $a$ in each of the \vplanet\ simulations is displayed in Figure \ref{fig:secular-evols}. Five of the six secular evolutions result in tidal heating that remains above the value for modern Earth for up to 7 Gyr. In the remaining simulation, Proxima b has $Q \simeq 16$. This result suggests that if Proxima b has $Q \sim 10$ (on the order of Earth), then the surface flux of tidal heating is not significant after the first few Gyr because the orbit has circularized. If, however, $Q \sim 10^2$, then tidal interactions render a first-order effect on the heat delivered to the world's surface over its entire history.}

\section{Discussion} \label{sec:discussion}

Our stability analysis has revealed key constraints in the orbital properties of Proxima d and b. In particular, eccentricities in excess of 0.5 are strongly disfavored, as are mutual inclinations between 95 and 142$^\circ$. \revone{We found cases in which the system is not Hill stable, and yet exhibits quasi-periodic motion. However, most quasi-periodic configurations of the system reside within the Hill stability boundary, as we should expect. Our analysis also identified regions where our simulations did not produce ejections, despite having crossing orbits initially. Although these configurations appear stable, a subset of simulations re-run for a longer integration time results in instabilities, suggesting that most of the region where $\beta < \beta_\text{crit}$ is in fact unstable.}


The separated islands of stable motion in Figure~\ref{fig:varying-ecc}, however, are very likely inaccessible. The large region of instability separating them from the low eccentricity ``mainland'' is \revone{probably} too large for the system to successfully cross. As shown in Figure~\ref{fig:all-ecc-evols}, instabilities can arise within a few centuries, which is short compared to the tidal damping timescale. We thus conclude that Proxima b and d could only have formed in the low eccentricity region of stability.

\revone{
Our simulations of secular orbital evolution in the system reveal that, beginning from the \revtwo{``stability mainland''} of parameter space determined with our $N$-body integrations, the tidal orbital evolution of Proxima b can easily \revtwo{continue} beyond 5 Gyr. In each case simulated with \vplanet, we see the eccentricity behaves as a damped harmonic oscillator, with secular amplitude decreasing rapidly with time. As such, we find no new constraint on the present-day eccentricity of this world, which has a value at $+1\sigma$ of 0.06 according to \citet{Faria2022}. This result suggests that tidal heating in Proxima b is likely ongoing, providing an additional source of energy to drive surface and atmospheric processes, perhaps including active biology. On the other hand, the increased volcanism could outgas water that is ultimately destroyed by high energy photons, resulting in a desiccated, and therefore uninhabitable world \citep{Luger2015}. Regardless, our research shows that tidal heating should not be ignored for planet b unless the upper limit of its eccentricity can be significantly reduced.
}

Models of the evolution and habitability of Proxima b can use our results to target their studies in the physically allowed regions we have identified in Figures \ref{fig:varying-ecc} and \ref{fig:varying-inc}. Climate studies \citep[e.g.][]{Turbet2016, delGenio2017,Colose21} could consider eccentricities for Proxima b up to 0.45, but should probably focus on the blue triangle in the lower left corner of Figure \ref{fig:varying-ecc}. Studies of the planetary interior could also consider the role of tidal heating over the same range \citep[see e.g.,][]{Henning2009, Barnes2013, DriscollBarnes2015, Barnes2016}. \revone{These results may also prove useful in dynamical modeling of the Proxima system from radial velocity data, providing limits on the maximum achievable orbital eccentricities and inclinations of planets d and b.}



\section*{Acknowledgments}
J.R.L. gratefully acknowledges partial support from a Mary Gates Research Scholarship. R.B. acknowledges support from NASA grant number 80NSSC20K0229 and the NASA Virtual Planetary Laboratory under grant number 80NSSC18K0829. The authors thank Gongjie Li\revone{, Hanno Rein, and Juliette Becker} for valuable insights and discussion\revone{, and thank the anonymous referee whose thoughtful recommendations substantially improved this manuscript}. This work was facilitated through the use of advanced computational, storage, and networking infrastructure provided by the Hyak supercomputer system and funded by the Student Technology Fund at the University of Washington.

\software{Rebound \citep{Rein2012}, VPLanet \citep{Barnes2020}, Astropy \citep{astropy2022}, Matplotlib \citep{Hunter2007}, Numpy \citep{Harris2020}, Pandas \citep{pandas-zenodo, pandas-paper}, Seaborn \citep{Waskom2021}.}

\bibliography{references}{}
\bibliographystyle{aasjournal}

\end{document}